\documentclass[sigconf,10pt,screen,authorversion]{acmart}

\usepackage[all]{nowidow}
\usepackage{xcolor}
\usepackage{subcaption}
\usepackage{hyperref}

\usepackage[capitalise,noabbrev]{cleveref}
\crefformat{section}{\S#2#1#3} 
\crefformat{subsection}{\S#2#1#3}
\crefformat{subsubsection}{\S#2#1#3}
\crefrangeformat{section}{\S#3#1#4 to~\S#5#2#6}
\crefrangeformat{subsection}{\S#3#1#4 to~\S#5#2#6}
\crefrangeformat{subsubsection}{\S#3#1#4 to~\S#5#2#6}

\newcommand{\komet}{\textsf{Komet}}

\copyrightyear{2024}
\acmYear{2024}
\setcopyright{rightsretained}
\acmConference[SoCC '24]{ACM Symposium on Cloud Computing}{November 20--22, 2024}{Redmond, WA, USA}
\acmBooktitle{ACM Symposium on Cloud Computing (SoCC '24), November 20--22, 2024, Redmond, WA, USA}
\acmDOI{10.1145/3698038.3698517}
\acmISBN{979-8-4007-1286-9/24/11}

\begin{document}

\title{\komet{}: A Serverless Platform for Low-Earth Orbit Edge Services}

\author{Tobias Pfandzelter}
\orcid{0000-0002-7868-8613}
\affiliation{%
    \institution{Technische Universität Berlin \& Einstein Center Digital Future}
    \city{Berlin}
    \country{Germany}}
\email{tp@3s.tu-berlin.de}

\author{David Bermbach}
\orcid{0000-0002-7524-3256}
\affiliation{%
    \institution{Technische Universität Berlin \& Einstein Center Digital Future}
    \city{Berlin}
    \country{Germany}}
\email{db@3s.tu-berlin.de}

\begin{abstract}
    Low-Earth orbit satellite networks can provide global broadband Internet access using constellations of thousands of satellites.
    Integrating edge computing resources in such networks can enable global low-latency access to compute services, supporting end users in rural areas, remote industrial applications, or the IoT.
    To achieve this, resources must be carefully allocated to various services from multiple tenants.
    Moreover, applications must navigate the dynamic nature of satellite networks, where orbital mechanics necessitate frequent client hand-offs.
    Therefore, managing applications on the low-Earth orbit edge will require the right platform abstractions.

    We introduce \komet{}, a serverless platform for low-Earth orbit edge computing.
    \komet{} integrates Func\-tion-as-a-Ser\-vice compute with data replication, enabling on-demand elastic edge resource allocation and frequent service migration against satellite orbital trajectories to keep services deployed in the same geographic region.
    We implement \komet{} as a proof-of-concept prototype and demonstrate how its abstractions can be used to build low-Earth orbit edge applications with high availability despite constant mobility.
    Further, we propose simple heuristics for service migration scheduling in different application scenarios and evaluate them in simulation based on our experiment traces, showing the trade-off between selecting an optimal satellite server at every instance and minimizing service migration frequency.
\end{abstract}

\maketitle

\section{Introduction}
\label{sec:introduction}

Sixth-generation mobile networks will be defined by an increasing focus on edge computing and the integration of non-terrestrial networks such as low-Earth orbit (LEO) satellite constellations~\cite{Bhattacherjee2018-vc,Handley2018-ay,lai2023starrynet}.
For this, edge computing brings compute and storage resources closer to clients, increasing both application service quality for users and enabling entirely new application domains with low-latency, high-bandwidth access to resources~\cite{shi2016edge,paper_bermbach2017_fog_vision,bonomi2012fog}.
At the same time, satellite networks comprising thousands of satellites in LEO will enable global access to high-bandwidth communications and Internet access~\cite{Bhattacherjee2018-vc,Bhattacherjee2019-jz,rahman20215g,Bhattacherjee2018-vc}.
While both edge computing and LEO networking have received considerable research and industry attention and are already commercially available, their combination, i.e., LEO edge computing, is still an emerging research field~\cite{Bhattacherjee2020-kr,paper_pfandzelter2021_LEO_serverless,paper_pfandzelter2022_celestial,wang2021tiansuan,Bhosale2020-aa,bhosale4don,lai2023starrynet}.

LEO edge computing refers to the integration of edge computing resources with LEO satellites, i.e., placing processors and storage on communication satellites.
Similar to terrestrial edge computing, the LEO edge could provide low-latency, high-bandwidth access to application services such as online collaborative drawing, multiplayer games, metaverses, web services, or the remote IoT, albeit for a global basis of subscribers.
A unique challenge of LEO edge computing is service orchestration:
Orbital dynamics dictate that LEO satellites must move at speeds in excess of 27,000km/h in relation to Earth, which results in frequent (on the order of 4--5 minutes) client handovers~\cite{Bhattacherjee2019-jz,Kassing2020-yc,Handley2018-ay}.
An edge service deployed on a LEO satellite will quickly be out of reach of a client, negating the benefits of proximity between clients and services.

The solution to this mobility is `virtual stationarity', where services transparently remain in client proximity by frequent service migration thus offsetting the satellites' movements~\cite{Bhattacherjee2020-kr}.
This requires that service migration is relatively cheap and can be completed without downtime.
Unfortunately, as we shall see in \cref{sec:migration}, this is not the case for the state-of-the-art service deployment model with container orchestration, e.g., using Kubernetes~\cite{Bhosale2020-aa}:
Frequently migrating stateful containers can lead to considerable service downtime during checkpoint, transfer, and restore operations.

Instead, we propose leveraging serverless abstractions for LEO edge computing.
By explicitly decoupling compute and storage services, we can deliver a large variety of LEO edge application services with virtual stationarity without downtime.
We make the following contributions:

\begin{itemize}
    \item We introduce \komet{}, a serverless platform for the LEO edge that combines an edge Function-as-a-Service (FaaS) compute platform with a distributed data management layer and automatically migrates services to provide virtual stationarity (\cref{sec:komet}).
    \item With \komet{}, we introduce heuristics for scheduling LEO edge services (\cref{sec:heuristics}).
    \item We demonstrate and evaluate \komet{} with a proof-of-concept prototype using typical edge computing applications on \emph{Celestial}~\cite{paper_pfandzelter2022_celestial} LEO edge testbeds.
    \item In simulations based on experiment traces, we investigate the trade-off between migration frequency and service level in scheduling LEO edge compute services (\cref{sec:simulations}).
\end{itemize}

We make our implementation artifacts available as open-source software.\footnote{\url{https://github.com/3s-rg/komet}}

\section{Background \& Related Work}
\label{sec:background}

To understand the challenges associated with service availability in LEO edge computing, we first give an overview of the characteristics of LEO satellite networks, the goals of edge computing in general, and the state of the art in LEO edge research.

\subsection{LEO Satellite Networks}

Large LEO satellite networks promise global low-latency, high-bandwidth Internet access~\cite{Bhattacherjee2018-vc,Handley2018-ay,mohan2024multifaceted,izhikevich2024democratizing}.
While traditional satellite Internet access using satellites in geostationary orbits has existed for decades, the high satellite altitude (more than 35,000km) incurs a high (more than 500ms round-trip time (RTT)) access delay.
LEO satellite networks use orbital altitudes of 500km to 1,000km, promising higher bandwidth (given reduced radio power requirements) and reduced access latency.

\begin{figure}
    \includegraphics[width=0.9\linewidth]{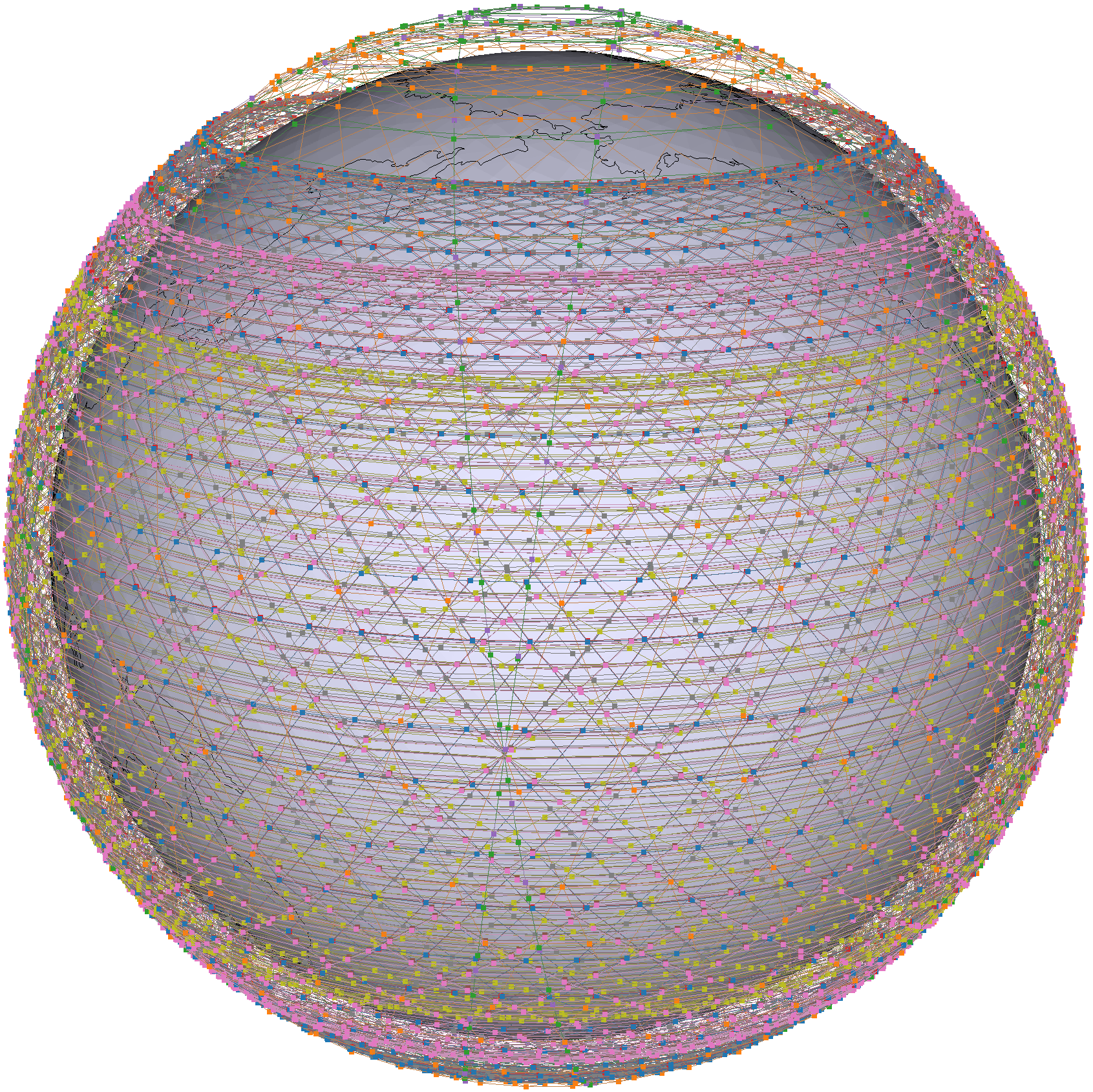}
    \caption{In its current deployment, the Starlink LEO satellite constellation comprises 4,409 satellites to achieve global coverage (screenshot from the Celestial emulation toolkit~\cite{paper_pfandzelter2022_celestial}). Lines between satellites indicate inter-satellite links (ISL).}
    \label{fig:starlink}
\end{figure}

Two characteristics of LEO satellites shape how LEO constellations achieve global network coverage:
First, a satellite that is close to Earth has a reduced cone of coverage, i.e., a single satellite can only serve a limited region on the ground.
Second, satellites in LEO move at speeds in excess of 27,000km/h, with an orbital period of one to two hours~\cite{Bhattacherjee2018-vc,Bhattacherjee2019-jz}.
Continuous global coverage by satellites thus requires constellations of hundreds or thousands of connected satellites.
As shown in \cref{fig:starlink}, SpaceX operates 4,409 Starlink satellites evenly spaced around Earth to achieve such global coverage~\cite{Kassing2020-yc}.
Other plans include Amazon Kuiper's proposed 3,236-satellite constellation and the 1,671-satellite Telesat constellation~\cite{Kassing2020-yc}.
To cover large distances, e.g., between an uplink station and a remote client, satellites use high-bandwidth inter-satellite links (ISL), and satellite networks essentially perform multi-hop routing between clients~\cite{Bhattacherjee2019-jz}.

\subsection{Edge Computing}

Edge computing extends the on-demand, elastic compute and storage resources of cloud computing throughout the access network, enabling application developers to provide their service in proximity to service consumers, e.g., IoT devices or mobile clients~\cite{shi2016edge,bonomi2012fog}.
While there are competing definitions of the term, we consider edge computing resources to be servers located at the access network, e.g., at a mobile radio tower, serving multiple clients with application services provided by multiple tenants~\cite{bonomi2012fog,paper_bermbach2017_fog_vision}.
Here, edge computing provides benefits in terms of service access latency and bandwidth as well as resiliency to backbone outages and contention.
Nevertheless, edge computing also introduces challenges, such as service management across geo-distributed nodes or resource allocation in environments that are more constrained than the cloud~\cite{paper_pfandzelter2020_tinyfaas,raith2023serverless}.

These challenges also apply when transferring the concept of edge computing to LEO satellite networks, i.e., the \emph{LEO edge}~\cite{Bhattacherjee2020-kr,bhosale4don,Bhosale2020-aa,cassara2022orbital,luglio2022performance,pfandzelter2023failure,paper_pfandzelter2021_LEO_serverless,paper_pfandzelter2021_LEO_CDN}.
As LEO networks are another kind of radio access network, albeit with access points on satellites in space, it can be envisioned that these access points may also provide compute resources that can be used by application providers to host services for network clients.
Here, resource management and especially resource sharing among tenants become more pressing concerns given the power and weight limits of satellites.
While research has shown that equipping network satellites with sufficient edge compute resources is possible~\cite{Bhattacherjee2020-kr}, it becomes paramount that any available resources are allocated as efficiently as possible.
Further, there is the challenge of node mobility, as LEO satellites frequently (every 4--5 minutes) move out of visibility of clients, requiring service migration to provide continuous coverage~\cite{Bhosale2020-aa,mohan2024multifaceted}.
As no deployed LEO edge systems are currently available, researchers use virtual LEO edge simulators, e.g., \emph{SatEdgeSim}, or emulators, e.g., \emph{Celestial}~\cite{paper_pfandzelter2022_celestial} and \emph{StarryNet}~\cite{lai2023starrynet}, to investigate the behavior of LEO edge applications.

\subsection{Serverless Computing}

One promising approach to manage application services by multiple tenants efficiently while still providing high levels of service isolation is serverless computing~\cite{hendrickson2016serverless,jonas2019cloud}.
In serverless computing, operational concerns are shifted to compute platform providers, while application developers focus on business logic.
Allocating underlying infrastructure, managing the core software stack, and elastically scaling services are all performed by the platform.
While this has obvious benefits for application developers, in the context of edge computing this shift of operational concerns to a shared underlying platform can also help achieve more efficient resource allocation~\cite{aslanpour2021serverless,paper_pfandzelter2020_tinyfaas,raith2023serverless}.

Function-as-a-Service (FaaS) as the most prominent serverless programming model allows developers to design and deploy applications as collections of small, loosely coupled \emph{functions}~\cite{hendrickson2016serverless,jonas2019cloud}.
Each function is invoked with a single message or event, has the ability to invoke further functions or interact with external services such as data stores, and (optionally) responds with a return value.
FaaS functions run on a FaaS platform that elastically scales function handlers (that run the function) in response to incoming events, including scaling to zero when no computation is necessary.
Functions can share their underlying runtimes, e.g., multiple functions written in the Python programming language can share a hardware and software stack, making them lightweight and easy to distribute over the network.

These attributes make the FaaS programming and execution model a good fit for edge computing~\cite{paper_pfandzelter2020_tinyfaas,raith2023serverless}:
Sharing a hardware and software stack means that limited edge resources can be shared to a high degree and elastically scaling function resources (by creating and destroying handlers) in response to demand enables a fine-grained allocation of these resources over time.

To build stateful applications on top of stateless FaaS, (serverless) data management and synchronization services can be integrated~\cite{10.1145/3485510,9799194,10.1145/3517206.3526275,pfandzelter2023enoki,10.1145/3472883.3486974,hetzel2021muactor}.
While trivial in the cloud, this poses challenges for distributed edge deployments where function instances in different geographic locations need to access the same data, introducing a trade-off between the goal of local function execution and the need to synchronize data access.
While still an area of active research, current proposals use, e.g., request routing to execute functions near their data dependencies~\cite{10.1145/3517206.3526275,9799194} or commutative replicated data types (CRDTs) to enable distributed concurrent execution without synchronization~\cite{shapiro2011comprehensive,jeffery2021rearchitecting,pfandzelter2023enoki}.

\subsection{Related Work}

Bhattacherjee et al.~\cite{Bhattacherjee2020-kr} propose the concept of serving edge applications from LEO satellites using a concept they call `virtual stationarity'.
Virtual stationarity provides an edge service from the same geographic location despite the highly dynamic LEO network infrastructure by handing off application state along with the hand-offs of clients, i.e., as satellites move along their orbits, services are moved in the opposite direction.
Further, the authors show two methods for server selection:
A \emph{MinMax} approach selects the closest server to a client at every instant, resulting in frequent changes.
An alternative \emph{Sticky} approach aims to minimize state transfers by predicting which satellite with a reasonable distance to the client (within 10\% of the optimum) will be in view of the client for the longest duration and has the smallest hand-off latency, thus trading off a small increase in access latency for less frequent hand-offs and hand-offs with low latency.
While these approaches are interesting theoretical mechanisms for managing LEO edge services, the authors provide no platform abstraction into which services can be deployed.

Existing work on LEO edge application platforms focuses largely on container orchestration, similarly to research on terrestrial edge computing.
Bhosale et al.~\cite{Bhosale2020-aa} present \emph{Krios}, an extension of Kubernetes for LEO satellite edge computing.
Kubernetes orchestration is reactive, i.e., the scheduler waits for an issue or error (possibly along with timeouts or retries) before deciding to reschedule a service.
In the context of LEO edge computing, the authors find that this leads to unnecessary downtime when a satellite that provides an application service for a client leaves the visibility of that client, as Kubernetes does not natively understand that the edge server can move away from a client.
Krios extends Kubernetes with proactive scheduling that takes satellite orbit models into account:
Krios will automatically schedule an application hand-off shortly before a satellite loses connection to a client, achieving continuous virtual stationarity.
As we discuss and show in \cref{sec:migration}, however, using container scheduling for LEO edge computing requires that container services are stateless, as migrating stateful containers comes with the cost of service downtime.
This could only be mitigated by explicitly designing the containerized service to externalize state or to gracefully handle concurrent instantiation at different satellite nodes without inconsistent data.
While this is indeed possible, it would expose the challenges of LEO edge computing to the application developers and essentially negate any developer benefits of platform abstractions.

C.~Wang et al.~\cite{wang2023satellite} and S.~Wang et al.~\cite{wang2024first} propose a \emph{cloud-native} satellite architecture.
This architecture uses cloud-native technologies such as Docker, Kubernetes, and edge-cloud networking on top of virtualized computing and storage to build flexible satellites.
The authors have successfully implemented this architecture on the \emph{BUPT-1} satellite that is part of the planned \emph{Tiansuan} constellation~\cite{wang2021tiansuan,xing2024deciphering}, showing that containerized computing in LEO is feasible.
This architecture targets isolated satellites or small satellite clusters rather than constellations of thousands of networked satellites, however, making their use-case and requirements different from the ones we discuss in this paper.

In previous work~\cite{paper_pfandzelter2021_LEO_serverless}, we proposed using FaaS as a LEO edge execution paradigm.
The combination of stateless functions and an external data management system supports migration of services against the orbital movement of LEO satellites that is transparent to clients, who experience continuous service availability, and application developers, who have explicit constraints on how to manage state in their applications by the programming model.
We have hitherto not provided an implementation or integrated architecture for this approach and did not evaluate its feasibility experimentally before this work.

These works all build on a strong foundation of research on orchestration and scheduling in terrestrial edge computing including work on edge container orchestration~\cite{bartolomeo2023oakestra,rausch2021optimized,brogi2020place,jeffery2021rearchitecting}, serverless edge computing~\cite{raith2023serverless,russo2023serverledge,aslanpour2021serverless,xie2021serverless}, and stateful edge-to-cloud computing~\cite{9799194,cheng2019fog,10.1145/3517206.3526275,pfandzelter2023enoki}.
The mobility of LEO edge satellite servers, however, means that we cannot directly use these existing approaches for LEO edge computing~\cite{Bhattacherjee2020-kr,paper_pfandzelter2021_LEO_serverless}.

\section{Downtime in LEO Edge Container Migration}
\label{sec:migration}

Using container orchestration for LEO edge services, such as in \emph{Krios}~\cite{Bhosale2020-aa}, requires service migration to achieve virtual stationarity.
To migrate a (stateful) container, it first needs to be checkpointed, i.e., the container is stopped and its memory content is written to disk.
Second, that checkpoint needs to be transferred to the new execution location, e.g., the next satellite a client will be handed off to.
Third, the container can be restored from the checkpoint on the new location.
We argue that this process, which has to be repeated continuously throughout the service's lifetime, leads to considerable service downtime.

Consider a simple example:
A client in Redmond, WA, USA, accesses an in-memory store on a LEO satellite in a single orbital plane of the phase \textrm{I} Starlink constellation (550km altitude, 22 satellites per orbital plane, i.e., 16.4° spacing between nodes~\cite{Kassing2020-yc}).
The mean contact length between the ground station and a satellite is 260 seconds, i.e., for an optimal latency between client and server, the container that hosts this service must be migrated every 260 seconds.

We implement this scenario on top of the Celestial LEO edge emulation toolkit~\cite{paper_pfandzelter2022_celestial} using a \emph{Redis} Docker container based on the lightweight \texttt{redis:alpine} image.
We use the \emph{Checkpoint/Restore In Userspace} (CRIU) software~\cite{criu} in combination with the \emph{Podman} container manager~\cite{podman} for check\-point-re\-store operations.
Our client loads data into the container's memory through the Redis API and continuously reads single values.
When the client is handed off to a new satellite, we checkpoint the container on the source satellite, migrate this checkpoint, and restore the container.
We measure checkpoint, transfer, and restore time for our service for different data sizes between 0 and 1,000MB.
The ground station and each satellite server have 4 vCPUs and 8GB of memory and run Alpine Linux 3.18, we assume 10Gbps bandwidth for ISLs and ground-to-satellite links.
Our host server has two 12-core Intel\texttrademark{} Xeon\texttrademark{} Silver 4310 2.10GHz CPUs, 64GB memory, SSD storage, and runs Ubuntu 22.04 LTS.
We run each experiment for 15 minutes and repeat it ten times.

\begin{figure}
    \centering
    \includegraphics[width=\linewidth]{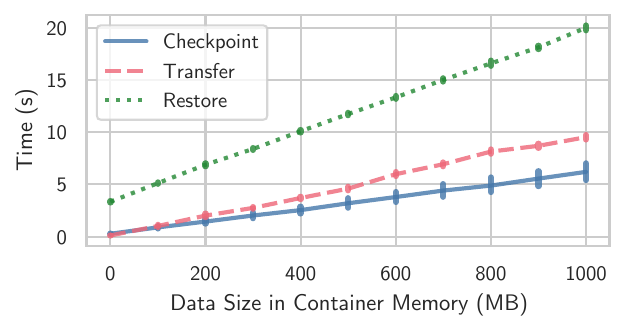}
    \caption{Container checkpoint, transfer, and restore times in our container migration experiment. It is obvious that the total migration time for a stateful container grows linearly with the amount of memory the container uses.}
    \label{fig:migration_time}
\end{figure}

Unsurprisingly, the results in \cref{fig:migration_time} show a linear correlation between the size of data in container memory and the migration time.
In our experiments, there is a minimum mean 3.82s migration delay (dominated by a mean 3.39s restore time) even without any data in the container's memory, simply to restore the started and configured Redis service.
For 1,000MB data, this migration delay grows to a mean 35.7s.
With a service migration every 260s to mitigate the effect of satellite orbital movement, this would lead to a considerable $\sim$14\% downtime.
Note that this includes only application state, while the Redis base image already exists at each host.
We expect similar replication costs if full virtual machines, microVMs, or unikernel VMs were used instead of containers, as process transfer costs remain.

While these results may not be representative of all hardware configurations and there are indeed advances in the efficiency of container migration~\cite{10.1145/3132211.3134460,9799256,10.1145/3301418.3313947}, it is clear that such an approach that is transparent to application developers, i.e., the software is not adapted for frequent migration, is \emph{not} transparent for service clients.

\section{\komet{} Architecture}
\label{sec:komet}

To address the complexity of building LEO edge applications that can be transparently migrated, we propose \komet{}, a serverless platform for the LEO edge.
\komet{} is built on the intuition that decoupling state from compute enables transparent live migration of services and that the semantics of FaaS are a well established approach for such decoupling.
The core technique that enables this is the concurrent deployment of two function instances backed by replicated data during the migration of a service.

\subsection{Application Runtime}

\begin{figure}
    \centering
    \includegraphics[width=\linewidth]{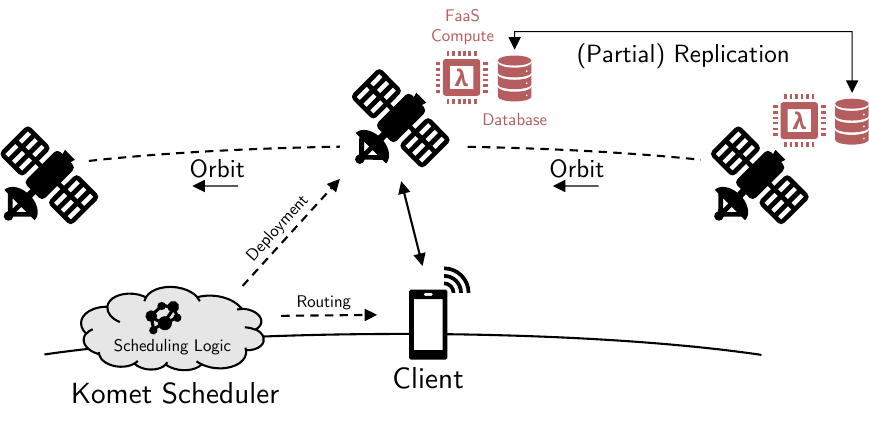}
    \caption{
        \komet{} comprises a per-satellite FaaS compute platform and a replicated data store, hosting replicas of the application service on multiple satellites.
    }
    \label{fig:architecture}
\end{figure}

\komet{} thus comprises two components on each satellite node:
A single-node FaaS platform that supports elastically running application services as FaaS functions and a data replication component that provides the stateful backend for those services.
We show the architecture of \komet{} in \cref{fig:architecture}.
In \komet{}, each satellite server hosts a FaaS compute platform and a database system.
\komet{} deploys FaaS functions on a satellite server near the client that wishes to invoke them.
Alongside the function handlers, each function has its own data pool in the database system to keep state across function invocations.
This data can be replicated during migrations to ensure that a client can seamlessly switch from one satellite service to the next.
As the function code itself is stateless, replicating only the data is sufficient for the service to transparently appear as a monolithic entity rather than a distributed application.
This has the additional benefit that only application state has to be replicated rather than the entire software stack, as with container migration.
The high-bandwidth, low-latency ISLs between satellites make data replication easily possible.

\subsection{Migration}

\begin{figure*}
    \centering
    \begin{subfigure}[t]{0.23\linewidth}
        \includegraphics[width=0.99\linewidth]{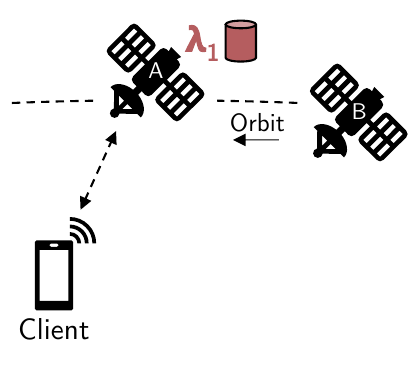}
        \caption{Client accesses service $\lambda_1$ at satellite A}
        \label{fig:migration:1}
    \end{subfigure}%
    \hfill
    \begin{subfigure}[t]{0.23\linewidth}
        \includegraphics[width=0.99\linewidth]{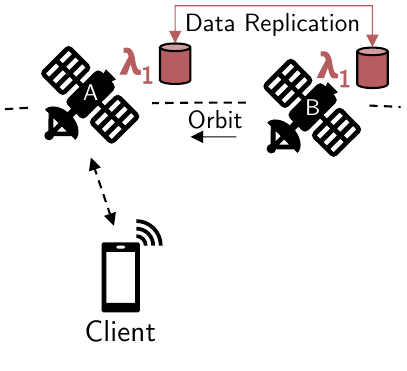}
        \caption{Service $\lambda_1$ is proactively replicated to satellite B}
        \label{fig:migration:2}
    \end{subfigure}%
    \hfill
    \begin{subfigure}[t]{0.23\linewidth}
        \includegraphics[width=0.99\linewidth]{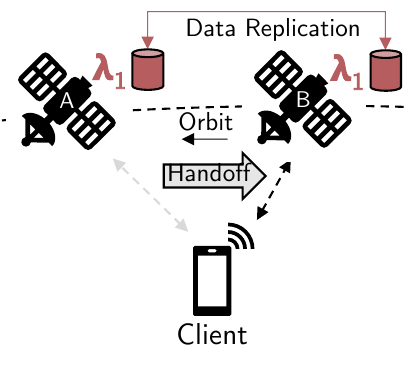}
        \caption{As client uplink is handed off to satellite B, client transparently continues to access $\lambda_1$ at client's new satellite B}
        \label{fig:migration:3}
    \end{subfigure}%
    \hfill
    \begin{subfigure}[t]{0.23\linewidth}
        \includegraphics[width=0.99\linewidth]{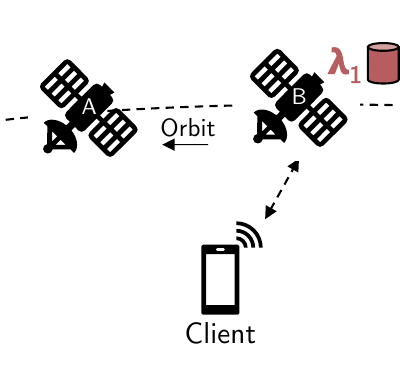}
        \caption{The copy of $\lambda_1$ can be removed from satellite A when it leaves client visibility}
        \label{fig:migration:4}
    \end{subfigure}%
    \caption{Service Migration in \komet{}}
    \label{fig:migration}
\end{figure*}

Data replication in \komet{} is also used to transparently migrate the service against the satellites' orbital movement, as shown in \cref{fig:migration}.
By default, a client expects the edge service it uses to be located on a nearby satellite server (\cref{fig:migration:1}).
As the satellite network evolves, i.e., the satellites follow their orbital paths, \komet{} proactively replicates the edge service to the selected next satellite (\cref{fig:migration:2}).
Here, only the state in the database is replicated, while the actual application processes are started anew, requiring no expensive process checkpointing.
Note that the highly predictable behavior of satellite orbits makes such calculations trivial and accurate.
When the client connection is handed off to the next satellite, the client can continue to access its edge service at the new location, without being aware of service migration occurring in the background (\cref{fig:migration:3}).
As soon as the original satellite serving the client is out of view, the service copy at that satellite can be deleted to reduce the replication costs (\cref{fig:migration:4}).
Finally, the process can be repeated for the next migration.

\subsection{Scheduler}

The migration of LEO edge services requires some coordinating entity that deploys \komet{} function replicas and synchronizes hand-offs between satellite service and client.
We propose a per-application, centralized (e.g., cloud-based) \komet{} scheduler.
This scheduler considers the global network state, i.e., clients, deployed services, network connections, and predicted network evolution based on satellite trajectories, and schedules service deployments accordingly.
Specifically, the scheduler is responsible for (i)~(proactively) deploying functions and data replicas to satellite servers, (ii)~informing clients when they should connect to a new satellite server, and (iii) removing functions and data replicas from servers when they are no longer used.
While a central scheduler may be less scalable than a distributed, on-satellite scheduling component, it is easier to reason about, and we believe that it does not impact service quality, as scheduling is asynchronous and not on the critical path of client requests to edge services.
Furthermore, satellite movement is so predictable that it would in fact be possible to schedule (and distribute) a migration plan that covers several days with at-runtime scheduling only sending minor live adaptations and synchronizing all clients and servers to perform migration at roughly the same time.
This would effectively remove the single point of failure, simply by having extensive time for recovery in case of failures.
Note also that centralized scheduling is not uncommon in satellite networks: SpaceX is believed to use a central scheduler for its entire Starlink network, updating ground stations and routes every 15s~\cite{mohan2024multifaceted,izhikevich2024democratizing}.
Still, we plan to investigate a distributed scheduler that is collocated with the edge server \komet{} platform in future work, as we discuss in \cref{sec:discussion}.

\section{Scheduling Heuristics}
\label{sec:heuristics}

The serverless abstractions in \komet{} provide flexibility in service migration and can be used with a variety of scheduling approaches.
Finding an optimal service migration schedule depends on many factors beyond service access latency and is thus not the aim of this paper (we discuss some avenues for further research on this in \cref{sec:discussion}).
Rather than finding an optimal schedule, we propose simple heuristics for service scheduling that serve as a starting point for LEO edge service scheduling with \komet{}.
Similarly to the \emph{Sticky} approach proposed by Bhattacherjee et al.~\cite{Bhattacherjee2020-kr}, we are interested in trading some service latency (being only within, say, 10\% of the optimum) for fewer service migrations.
While service migrations in \komet{} are seamless, replicating data across the network can still be costly in terms of bandwidth and should be avoided where possible.

We propose heuristics for three scenarios: a single client accessing a single service instance (\emph{one-to-one}), multiple clients sharing a single service instance (\emph{many-to-one}), and multiple clients sharing multiple service instances (\emph{many-to-many}).

\subsection{One-to-One}

For a single client, we can proactively migrate the LEO edge service instance by calculating the trajectory of the satellite network, which is not compute-intensive~\cite{Kempton2021-lw,Handley2018-ay}.
This allows us to determine the network distance between the client and all available satellite servers in the future.
Our heuristic first simply selects the closest satellite but afterwards only switches servers if a new server is at least 10\% better in terms of network distance compared to the currently selected satellite.
The 10\% threshold is adapted from the existing \emph{Sticky} approach as an arbitrary point along the trade-off between service latency and migration frequency, and we measure the impact of this threshold (compared to others) in \cref{sec:simulations}.
Further, we also measure service migration delay (time to deploy the necessary functions and data replica to the new satellite server) and initiate this deployment with sufficient lead-time, ensuring that the service is ready when the hand-off needs to be performed.

\subsection{Many-to-One}

For many clients sharing a single server, we can extend our heuristic to take multiple network distances into account by aggregating them.
We use the root-mean-square for this aggregation as it is more sensitive to outliers than the mean, ensuring that all clients have comparable service access latency.
This can similarly be calculated ahead-of-time, and we again only initiate hand-offs if the score for a new satellite is at least 10\% better than the currently selected one.

\subsection{Many-to-Many}

FaaS functions backed by replicated data in \komet{} also opens up the possibility of a many-to-many deployment: Multiple clients share multiple instances of the service that are all replicated.
Each client can access a service copy at a satellite near itself, while the underlying state is replicated, essentially providing identical services at different locations.
The difference to the many-to-one deployment is the trade-off between service latency and data staleness.

Selecting multiple servers requires a more advanced heuristic as there is now the additional cost of each additional replica (in terms of, e.g., bandwidth for replication).
We follow a similar approach using pre-calculated satellite trajectories.
For each client, we first calculate the set of satellite servers that are within 10\% network distance of the closest satellite to that client.
We then calculate a \emph{hitting set}, i.e., the smallest set of satellites that contains at least one of each client's closest satellites.
While this is theoretically NP-hard, note that there (i)~exist sufficient approximations~\cite{10.1145/285055.285059}, (ii)~the number of satellites is limited, and (iii)~this calculation is not on the critical path in live scheduling, as trajectories can be pre-calculated.
The number of satellites required to host the service may thus also change over time.
For each newly selected satellite, we select the closest (in terms of network distance) satellite currently running the service as a replication source.
This means that multiple satellites could receive data from the same source satellite and that some satellites may not migrate their local state further.
Further, note that this heuristic only optimizes for service latency and does not take into account the resource requirements for serving multiple clients.
We consider this out of scope for this work but discuss avenues for future research on scheduling in \cref{sec:discussion}.
\section{Prototype Implementation}

\begin{figure}
    \centering
    \includegraphics[width=\linewidth]{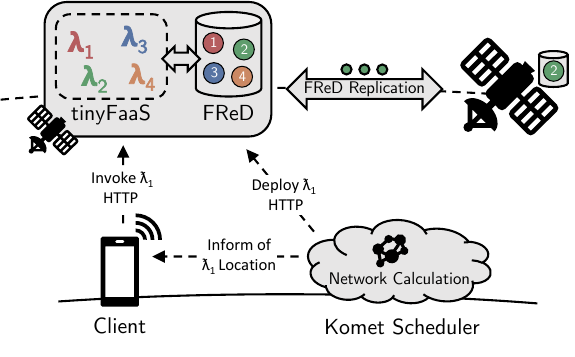}
    \caption{Our \komet{} prototype combines the tinyFaaS lightweight edge FaaS platform and FReD edge data management middleware with a central scheduling component. Functions are backed by replicated data and can be invoked by clients through an HTTP endpoint.}
    \label{fig:implementation}
\end{figure}

To evaluate \komet{}, we implement a proof-of-concept prototype.
We combine existing research prototypes in the field of serverless edge computing with a novel LEO-edge-focused scheduling component that handles proactive replication.
Note that the goal of our prototype is not to provide a full-fledged software system with the best possible performance but rather to illustrate and evaluate the \komet{} concept.
We show an overview of our implementation in \cref{fig:implementation}.

We use the \emph{tinyFaaS}~\cite{paper_pfandzelter2020_tinyfaas} lightweight FaaS platform that can run Python functions on a single Linux host.
tinyFaaS exposes an HTTP endpoint to invoke functions and a separate HTTP endpoint to upload and remove functions.
Function handlers are isolated using Docker containers.
While this does not necessarily meet the security requirements of all services, we consider the issue of efficient isolation for multi-tenant FaaS systems orthogonal to our work.
We select tinyFaaS specifically for its low overhead and resource footprint, but the \komet{} architecture could also be implemented with other FaaS platforms such as \emph{OpenFaaS}~\cite{openfaas}, \emph{Knative}~\cite{knative}, or \emph{nuclio}~\cite{nuclio}.

Further, we rely on the \emph{FReD}~\cite{pfandzelter2023fred} data management platform for geo-distributed edge-to-cloud environments.
FReD manages multiple independent data pools, called \emph{keygroups}, across a distributed set of servers, called \emph{nodes}.
Specifically, clients can dynamically specify to which nodes a keygroup should be replicated, and FReD will ensure data replication for that keygroup with client-centric consistency guarantees.
While FReD only supports key-value data with a simple CRUD interface, this is sufficient for our proof-of-concept prototype.
Again, \komet{} could also be implemented with alternative data management tools that offer application-controlled replica placement~\cite{poster_hasenburg2020_towards_fbase,techreport_hasenburg2019_fbase}.

The prototype of our \komet{} scheduler is implemented in Python.
Clients connect to this scheduler using a WebSocket and receive new server locations as soon as a service is migrated.

\section{Demonstration \& Experimental Evaluation}
\label{sec:demonstration}

We evaluate \komet{} using our prototype implementation on the Celestial LEO edge emulator~\cite{paper_pfandzelter2022_celestial}.
Celestial emulates LEO satellite and ground station servers using Firecracker microVMs~\cite{agache2020firecracker} and adjusts network parameters such as latency and bandwidth according to a simulated LEO constellation.
We deploy three different applications on our \komet{} prototype: a simple read-write cache for a single client, evaluating the feasibility of \komet{} (\cref{sec:demonstration:simple}); an IoT application that implements a shared service for multiple clients (\cref{sec:demonstration:iot}); and a content delivery network demonstrating a multi-user application with a shared, distributed cache (\cref{sec:demonstration:cdn}).
As our intention with the \komet{} prototype is that it serves as a proof-of-concept of our approach, we focus on latency and migration performance in this evaluation, showing that \komet{} can maintain a consistent service level for different kinds of applications over time, despite the dynamic satellite network topology.

\subsection{Single-Client Cache}
\label{sec:demonstration:simple}

\begin{figure}
    \centering
    \includegraphics[width=0.6\linewidth]{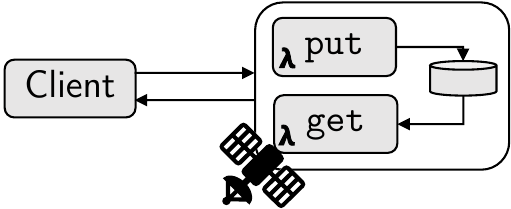}
    \caption{Single-client cache service: The application has a \texttt{get} and a \texttt{put} function that read and write data into the cache, respectively. The client alternates between calling these two functions every second.}
    \label{fig:simple}
\end{figure}

Our simple read-write cache follows the structure of our motivating example in \cref{sec:migration}:
We deploy our cache as two functions, as shown in \cref{fig:simple}.
The \texttt{put} function writes data into the backend data store, while the \texttt{get} function returns data for a key.
A single client in Redmond, WA, USA, accesses this cache, reading and writing a single data item every second.
We again use a single plane from the phase \textrm{I} Starlink constellation with 22 satellites evenly spaced at 550km altitude.
This implementation uses our one-to-one scheduling heuristic, and preliminary measurements have revealed a 20-second migration time (data replication and function instantiation) that our scheduler takes into account for proactive migration.
Our functions and scheduler are implemented in Python, while the client is a static Go binary.
Our ground station and each satellite server again have 4 vCPUs and 8GB of memory, with 10Gbps bandwidth for ISLs and ground-to-satellite links, and we use the same 24-core host server.
Our experiment runs for a total of 15 minutes.

\begin{figure}
    \centering
    \includegraphics[width=\linewidth]{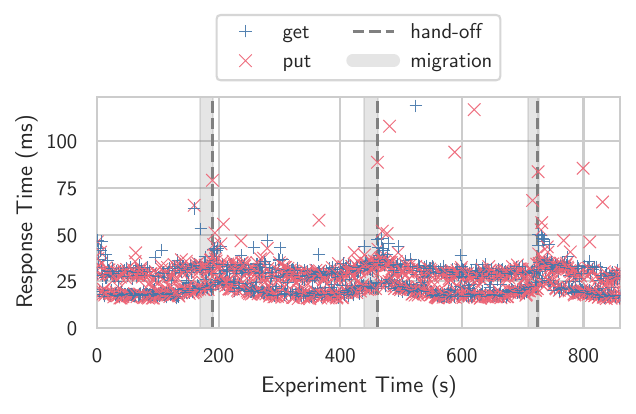}
    \caption{Request-response latency for each request during the single-client experiment along with migration and hand-off times. Results show a consistent client service level that follows orbit patterns, as the client's closest satellite passes over the client location.}
    \label{fig:simple_results}
\end{figure}

We show the request-response latency for each client request along with migration periods and hand-off events in \cref{fig:simple_results}.
As the service is deployed on the client's nearest satellite, request-response latency follows the familiar pattern of shrinking and growing as the satellite passes over the client (except for outliers, where the service is slow to respond).
The \komet{} orchestrator proactively initiates service migration before the client is handed off, as shown in the figure: During these migration periods, the service is replicated across the client's nearest satellite and the anticipated \emph{soon-to-be nearest} satellite.
In our experiments, this replication occurs 20.8s before the hand-off, on average.
In other words, the service is replicated for $\sim$2.4\% of our experiment time.
As soon as the hand-off is completed, the original replica is destroyed and only a single service replica on the client's then nearest satellite server remains.
As our results show, this hand-off technique leads to consistent request-response times without downtime.

\begin{figure}
    \centering
    \includegraphics[width=\linewidth]{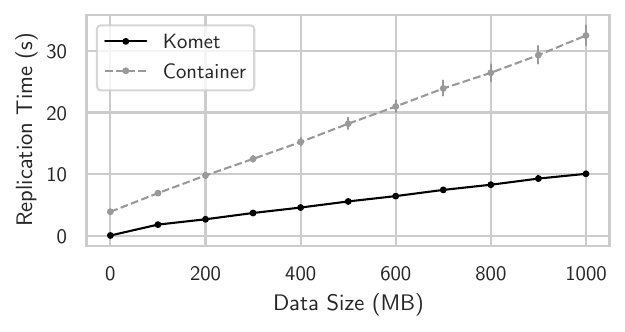}
    \caption{Time to replicate data to a new satellite server in our \komet{} implementation compared to total time for container migration (single client)}
    \label{fig:simple_migration}
\end{figure}

Naturally, the replication time in \komet{} also depends on the size of data in the data store.
To directly compare the performance of \komet{} with our container replication experiment in \cref{sec:migration}, we measure data replication times in our setup using between zero and 1,000 (in increments of 100) 1MB items.
The results in \cref{fig:simple_migration} show a similar, expected linear correlation between data size to replicate and replication times.
In \komet{}, however, this replication time does not lead to service downtime.
Further, the small minimum replication time of only 131ms to replicate the dataset with 0MB (compare 3.74s with containers) demonstrates the benefits of replicating only decoupled application state rather than the entire execution stack.

\subsection{Internet of Things}
\label{sec:demonstration:iot}

A more complex application of LEO edge computing is supporting remote IoT devices.
The \emph{National Oceanic and Atmospheric} (NOAA) \emph{Deep-ocean Assessment and Reporting of Tsunami} (DART)~\cite{Gonzalez1998-yo,Meinig2005-qq} ocean buoys are located throughout the Pacific Ocean and collect seismic measurements to detect tsunamis early.
As these buoys are far from any terrestrial network connection, they already use satellite networks to send data to a central monitoring station.
We have previously proposed equipping the satellite network that supports the DART buoys with compute resources to aggregate sensor data and perform ML inference to detect tsunami risks with low latency~\cite{paper_pfandzelter2022_celestial}.
Specifically, they have implemented a sample \emph{Long Short-Term Memory} (LSTM) ML model that runs on satellite servers.

\begin{figure}
    \centering
    \includegraphics[width=\linewidth]{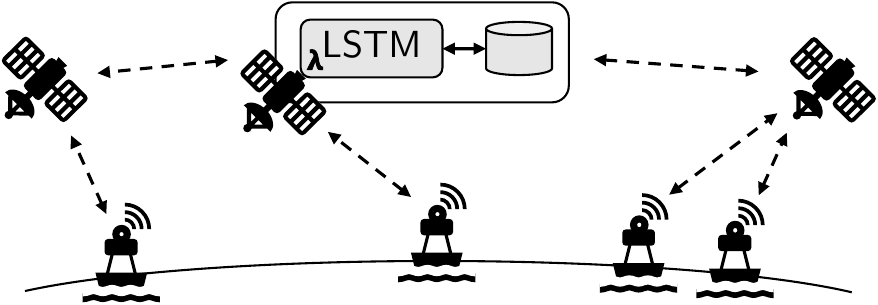}
    \caption{Our IoT application connects 30 DART seismic monitoring buoys with a shared LSTM service running on a satellite server. Current values for each sensor are stored in the backing database and ML inference is performed over all values.}
    \label{fig:iot}
\end{figure}

We use this existing implementation as a starting point for a serverless remote IoT application based on \komet{}.
As shown in \cref{fig:iot}, our application runs the existing LSTM model (implemented in TensorFlow) as a serverless function.
Remote buoys send their measurements to this function, which stores the latest value for each buoy in the backend data store.
Using these values, the LSTM model generates a risk factor and sends it as a response.
Multiple clients share the same service instance.

We use the locations of all 30 DART buoys in the North Pacific Ocean for our clients and the phase \textrm{I} Starlink constellation as our LEO network.
Buoys send a sensor measurement every second.
Our \komet{} scheduler is located on Fort Island, HI, USA (the location of the Pacific Tsunami Warning Center), although we do not expect this to impact our results.
This implementation uses the many-to-one scheduling heuristic.
Our functions and scheduler are again implemented in Python, while the client is a static Go binary.
All ground stations and satellite servers have 8 vCPUs and 8GB of memory, with 10Gbps bandwidth for ISLs and ground-to-satellite links.
Given the resource requirements, we perform this experiment on eight \texttt{n2-standard-16} Google Compute Engine VM instances with 16 vCPUs and 64GB of memory in the \texttt{europe-west3} (Frankfurt, Germany) region.
Our experiment runs for a total of 20 minutes plus two minutes of ramp-up time.

\begin{figure}
    \centering
    \includegraphics[width=\linewidth]{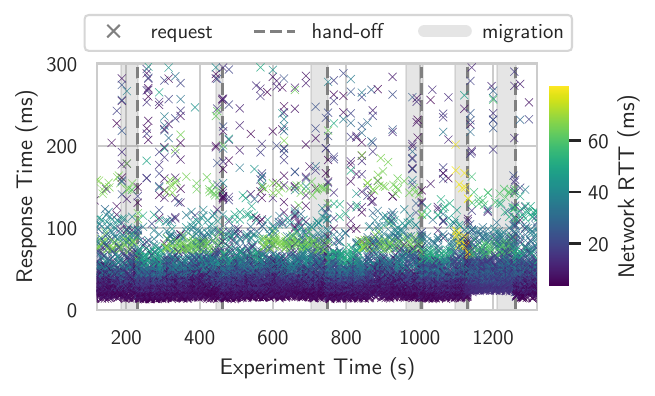}
    \caption{
        Response latency for each function invocation along with migration and hand-off times in the IoT experiment. Colors indicate calculated network round-trip times. Our heuristic leads to seven hand-offs during our 20-minute experiment. While there is some variance in the response latency, this shows how service response times remain consistent regardless of satellite movement.
    }
    \label{fig:iot_results}
\end{figure}

We show the response times for client requests in \cref{fig:iot_results}.
Additionally, we also show the expected network round-trip time from each client to the satellite server currently running the LSTM service.
While there are some outliers, we can see the correlation between request-response latency and network latency, i.e., response times are higher if the client is further away from the selected server.
Over the course of the 20-minute experiment, \komet{} migrates the service seven times.
The measured response times show that this keeps the service at a consistent service level for clients.

\subsection{Content Delivery Network}
\label{sec:demonstration:cdn}

A benefit of attaching replicated data storage to our LEO edge functions is that beyond sharing a single satellite server, clients can also share a LEO edge service that is distributed across multiple servers.
Consider content delivery networks (CDN), a further possible LEO edge application~\cite{Bhosale2020-aa,Bhattacherjee2020-kr}:
Clients use a nearby CDN caching server to request web pages or data, such as images or videos.
If the cache has a copy of the requested item, it is served to the client from this nearby cache.
If the cache does not have a copy, the item is first pulled from an origin location, then cached locally, and finally sent to the client.
While this has benefits for a single client, the benefit is even greater for multiple clients sharing a regional CDN caching layer, as clients in geographic proximity can exhibit similar request behavior.

\begin{figure}
    \centering
    \includegraphics[width=\linewidth]{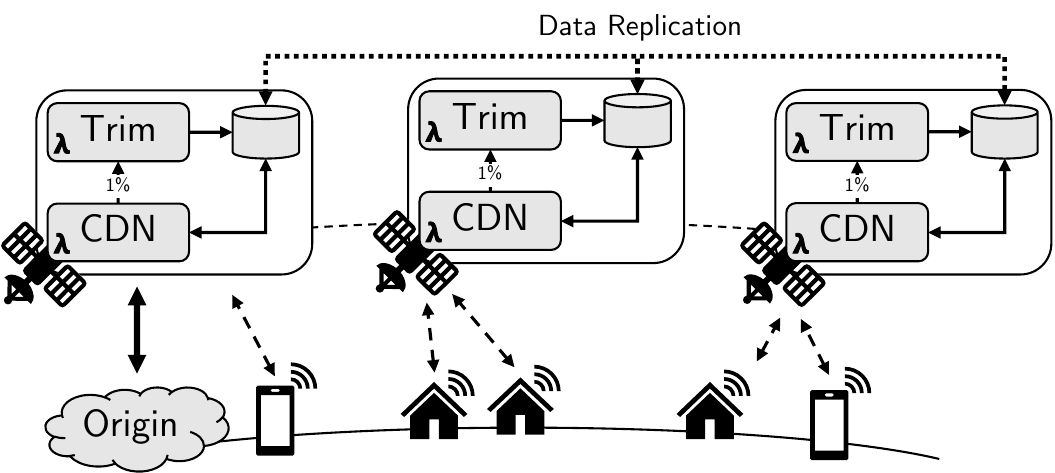}
    \caption{In the LEO edge CDN based on \komet{}, clients request data items from the CDN function deployed on their closest satellite. Data items are served from the replicated cache if available or first pulled from the origin location. A trimming function compacts the replicated cache periodically (randomly with a 1\% chance).}
    \label{fig:cdn}
\end{figure}

We can implement such a CDN cache on top of \komet{} as shown in \cref{fig:cdn}.
Multiple copies of the CDN function are distributed across satellites over a wider geographic area.
The CDN functions can use their local data store replicas to read and write data.
If a client accesses a file that is not in the store, any service replica can pull it from the origin location and store it.
Periodically (randomly with a 1\% chance in our implementation), a separate trimming function is called asynchronously to trim the cache to the most-recently used 1,000 items.

We use 50 client locations in the Northwest of the United States and an origin location in Umatilla County, OR, USA (site of an AWS data center in that region~\cite{awsoregon}).
Our data set is based on image request traces from the \emph{Wikimedia Mediacounts} data set~\cite{wiki-mediacounts}, from which clients request a data item every five seconds.
The scheduler in this experiment uses the many-to-many scheduling heuristic.
Our functions and scheduler are again implemented in Python, while the client is a static Go binary.
All ground stations and satellite servers have 8 vCPUs and 8GB of memory, with 10Gbps bandwidth for ISLs and ground-to-satellite links.
We perform this experiment on 16 \texttt{n2-standard-16} Google Compute Engine VM instances with 16 vCPUs and 64GB of memory in the \texttt{europe-west3} (Frankfurt, Germany) region.
Our experiment runs for a total of 20 minutes plus two minutes of ramp-up time.

\begin{figure}
    \centering
    \includegraphics[width=\linewidth]{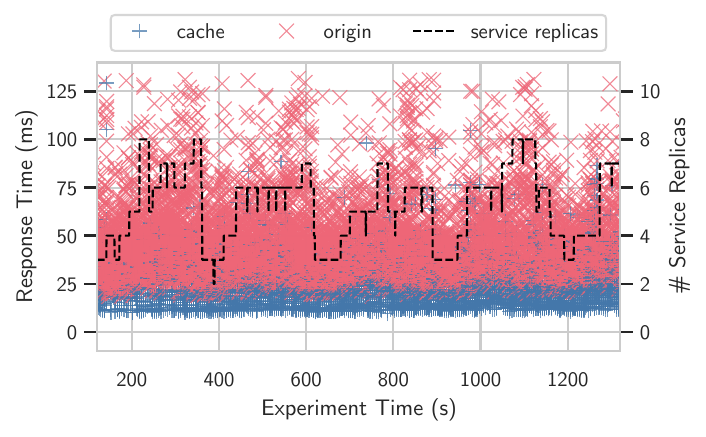}
    \caption{Response times for client requests in our CDN experiment. As expected, responses served from the CDN cache are faster than those pulled from the origin location. We also show the number of service replicas deployed by our scheduler over time.}
    \label{fig:cdn_results}
\end{figure}

We show the response times for client requests in \cref{fig:cdn_results}.
Unsurprisingly, responses served from the CDN cache are served with lower latency than those that first have to be pulled from the origin location.
On average, cached responses take 21.58ms while non-cached responses take 46.44ms.
In this workload, 34.8\% of requests can be served from cache, although only 1,000 (of a possible 542,112 in the dataset) are kept in the cache.

Interesting to observe is also the number of service replicas deployed over the course of the experiment:
Our scheduler deploys a maximum of eight replicas and a minimum of two replicas to serve all 50 clients, with a mean 5.3 replicas deployed.
This demonstrates an interesting dynamic in satellite networks, where a highly varying number of satellites is required to provide coverage for the same, static clients over time.

\section{Scheduling Simulations}
\label{sec:simulations}

We evaluate the effectiveness of our scheduling heuristics in simulations based on traces generated in our three experiments.
Here, we focus on evaluating the proposed heuristics in an isolated manner, without the additional effects of service migration, service performance, resource variability, and especially without being influenced by our particular \komet{} implementation.

\subsection{Single-Client (One-to-One)}

Using Celestial, we generate traces for the single-client cache for a client in Redmond, WA, USA, using the full first shell of the phase \textrm{I} Starlink, which comprises 1,584 satellites at 550km altitude.
These traces contain the network state at every second for a total duration of 20 minutes.
Using these traces, we simulate the behavior of different server selection strategies in \komet{}, noting the distance between the client and the selected satellite server as well as the duration between service hand-offs.
We compare the MinMax approach~\cite{Bhattacherjee2020-kr} and our one-to-one heuristic with different parameters: the default 10\% threshold for service migration, a 25\% threshold, and a threshold of 1ms, where a new satellite is only selected if it can provide service latency at least 1ms lower than the currently selected node.

\begin{figure}
    \centering
    \includegraphics[width=\linewidth]{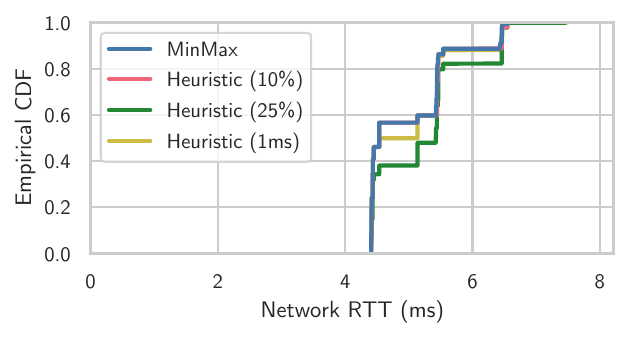}
    \caption{Network RTT between the single client and the satellite server the scheduling strategies select for that client. All strategies perform similarly, with a minimum mean 4.99ms RTT for MinMax and a maximum mean 5.22ms RTT for the \komet{} heuristic with a 25\% threshold.}
    \label{fig:distance_simple}
\end{figure}

\begin{figure}
    \centering
    \includegraphics[width=\linewidth]{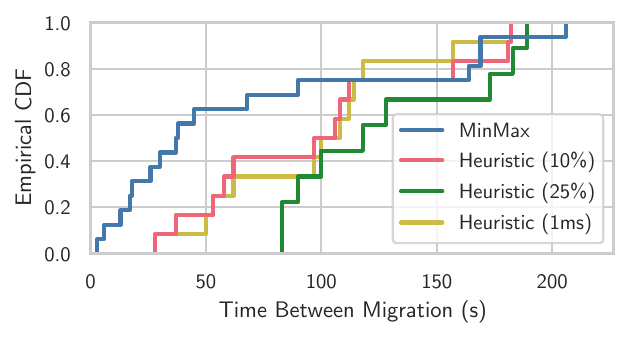}
    \caption{Time between service migrations differ in our single-client scenario, with a mean 68.7s duration for MinMax and 127.4s for the 25\% heuristic.}
    \label{fig:duration_simple}
\end{figure}

The network distance measurements for this simple experiment in \cref{fig:distance_simple} show similar results for all four heuristics, with a mean 4.98ms RTT for the MinMax strategy, 4.99ms RTT for our 10\% heuristic, 5.22ms for the 25\% heuristic, and 5.04ms for the 1ms heuristic.
Despite these nearly identical results, the duration between hand-offs shown in \cref{fig:duration_simple} show the different behaviors of our heuristics.
During the 20-minute trace, the MinMax strategy performs 16 service migrations, whereas the 10\% and 1ms heuristics perform twelve and the 25\% heuristic performs only nine.
On average, a service in the MinMax is active for only 68.7s, 98.4s in the 10\% and 1ms heuristic, and 127.4s in the 25\% heuristic.
This illustrates how our heuristic trades a marginal increase in RTT for significantly fewer of the costly service migrations.

\subsection{IoT (Many-to-One)}

To further investigate the trade-off between service distance increase and migration count, we use traces generated from the IoT experiment with 30 clients in the North Pacific.
Here, we again compare the MinMax strategy~\cite{Bhattacherjee2020-kr} (the lowest sum of client-satellite distance at every instance) to our heuristic with 10\% (default) and 25\% thresholds.
Additionally, we also simulate the Sticky strategy~\cite{Bhattacherjee2020-kr}.
Note that the Sticky strategy was intended only for a many-to-one scenario, and porting it to other evaluation scenarios would give limited comparability to existing work.

\begin{figure}
    \centering
    \includegraphics[width=\linewidth]{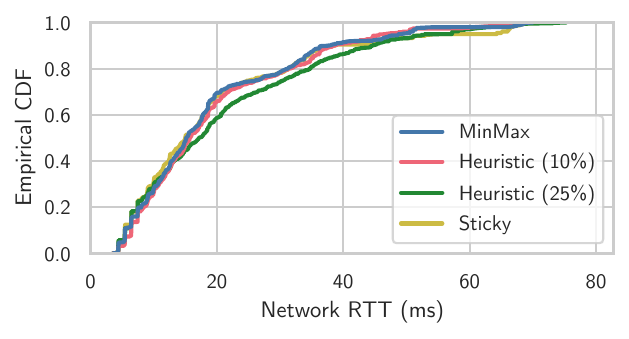}
    \caption{RTT measurements between clients and the service in our IoT scenario are similar regardless of the chosen strategy, with a mean 19.31ms, 19.78ms, 21.34ms, and 19.68ms RTT for MinMax, 10\% heuristic, 25\% heuristic, and Sticky, respectively.}
    \label{fig:distance_iot}
\end{figure}

\begin{figure}
    \centering
    \includegraphics[width=\linewidth]{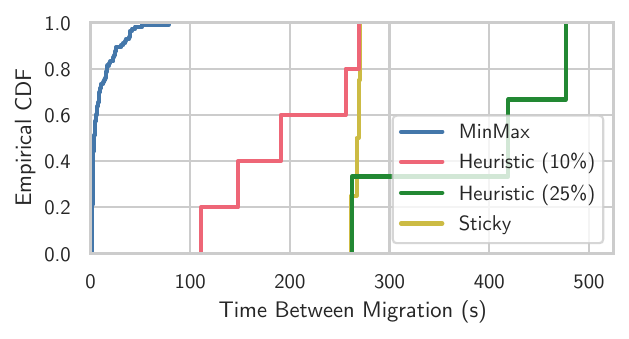}
    \caption{The server selection strategies differ in the amount of service migrations during the 20-minute trace, with MinMax performing 113 migrations in the IoT experiment (mean 10.33s between migrations) and our heuristics with 10\% and 25\% thresholds performing 5 and 3 migrations (mean 195s and 386s between migrations), respectively. Sticky performs 4 migrations with a mean 266.75s between migrations.}
    \label{fig:duration_iot}
\end{figure}

The distance measurements of our IoT experiments for the trace period of 20 minutes (\cref{fig:distance_iot}) again show how similar the MinMax and heuristic strategies perform.
The mean RTT is 19.31ms for MinMax, 19.78ms for the 10\% heuristic, and 19.68ms for Sticky.
Only the 25\% heuristic has a slightly higher mean RTT of 21.34ms.
More interesting here is the time between service migrations, shown in \cref{fig:duration_iot}.
During the 20-minute trace, the 10\% heuristic performs five service migrations, while the 25\% heuristic performs only three.
Compare this to MinMax, which requires a total of 113 service migrations for marginal benefits in service distance.
As a result, service instances in MinMax are active for only 10.33s on average before being migrated to the next satellite, compared to 195s (10\% heuristic) and 386s (25\% heuristic).
The Sticky heuristic performs similarly, requiring four migrations with a mean 266.75s between migrations.

\begin{figure}
    \centering
    \includegraphics[width=\linewidth]{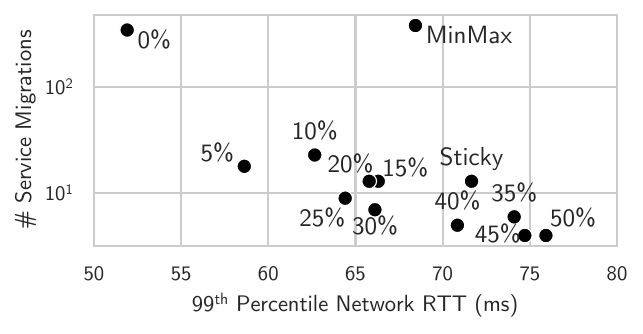}
    \caption{Increasing the threshold in our scheduling heuristic decreases the number of migrations while increasing the 99\textsuperscript{th} percentile latency, clearly showing a Pareto front in the trade-off between the number of migrations and the possible RTT.}
    \label{fig:pareto_iot}
\end{figure}

To further investigate this trade-off, we extend our trace generation to one hour for the IoT scenario.
We now simulate different thresholds for our heuristic, between 0\% and 50\% in five percentage point increments.
We show the achieved 99\textsuperscript{th} percentile RTT and the number of migrations during our one-hour trace in \cref{fig:pareto_iot}.
As expected, the MinMax and 0\% heuristic strategies perform the most migrations, at 348 over the course of one hour.
Note that the MinMax and 0\% heuristic only differ in terms of aggregation, with MinMax using the average distance from clients to servers and our heuristic using the root-mean-square to score potential satellites.
The result is a lower 99\textsuperscript{th} percentile latency as the root-mean-square is more fair to outliers (mean RTT is slightly higher at 19.75ms compared to 19.33ms).
Increasing the threshold in our scheduling heuristic decreases the number of migrations while increasing the 99\textsuperscript{th} percentile latency, clearly showing a Pareto front in the trade-off between the number of migrations and the possible RTT.
The results also show that our heuristic is not optimal in all cases, with the 5\% threshold leading to less hand-offs (18 compared to 23) and 99\textsuperscript{th} percentile RTT (58.62ms compared to 62.66ms) than the 10\% threshold.
A possible reason is that the 5\% heuristic makes a satellite selection that turns out to be more stable than the other heuristic as a result of the dynamics of the satellite network.

\subsection{CDN (Many-to-Many)}

We simulate our many-to-many heuristic using traces from our CDN example with 50 clients in Northwest USA.
We again compare the default 10\% heuristic from \komet{} with a 25\% threshold, 1ms threshold (select minimum set of servers to cover all clients within 1ms of their optimum distance).

\begin{figure}
    \centering
    \includegraphics[width=\linewidth]{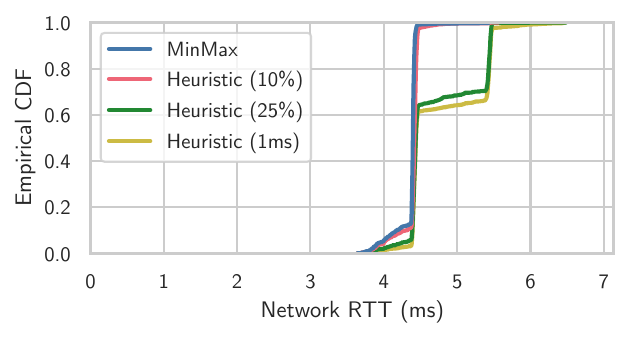}
    \caption{The network distances between clients and their selected satellite servers in the CDN use-case are similar between server selection strategies, between mean 4.36ms for MinMax and 4.79ms for the 1ms heuristic.}
    \label{fig:distance_cdn}
\end{figure}

\begin{figure}
    \centering
    \includegraphics[width=\linewidth]{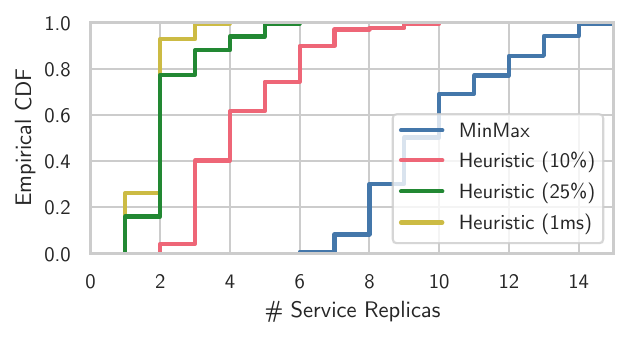}
    \caption{The CDF shows the number of concurrent service replicas each strategy needs in every second of the simulation, a lower number is better (less replication necessary).
        The strategies require different numbers of service replicas, with MinMax requiring a mean 9.84 and the 1ms heuristic requiring only 1.82.}
    \label{fig:num_satellites_cdn}
\end{figure}

As shown in \cref{fig:distance_cdn}, the distance measurements from our clients to their nearest service instance follow a trend that is similar to our single-client simulation:
MinMax and our 10\% heuristic achieve a comparable mean 4.36ms and 4.39ms RTT, respectively, with a slightly higher 4.73ms and 4.79ms for the 25\% and 1ms heuristics.
For this scenario, calculating the time between migrations is not possible, as there are multiple concurrent service instances.
Instead, we are interested in the number of such service instances, which we show in \cref{fig:num_satellites_cdn}.
Despite the highly similar RTT the different strategies achieve, they require different numbers of service replicas.
The MinMax chooses up to 15 service replicas (mean 9.84) throughout the 20-minute trace, while the 10\% heuristic requires only a maximum 10 (mean 4.36).
The more relaxed 25\% and 1ms thresholds again lead to less service replicas, with a maximum six (mean 2.25) and four (mean 1.82), respectively.
This demonstrates a similar trade-off to the one between service distance and migration frequency also in scenarios with multiple service replicas, i.e., this also affects utilization of compute resources in the satellite constellation.

\section{Discussion \& Future Work}
\label{sec:discussion}

We have introduced \komet{}, a serverless framework for LEO edge services that handles service migration seamlessly.
This section investigates limitations of our current architecture and derives opportunities for future work.

\subsubsection*{Limitations of the FaaS Programming Model}

In our evaluation, we have shown how the FaaS programming model we use in \komet{} can be employed to build a variety of LEO edge applications.
Beyond that, FaaS has been demonstrated to be a particularly good fit for edge applications in general~\cite{raith2023serverless,russo2023serverledge,aslanpour2021serverless,xie2021serverless}, especially with the addition of a data management backend~\cite{9799194,cheng2019fog,10.1145/3517206.3526275,pfandzelter2023enoki}.
Nevertheless, the FaaS programming model has limitations:
For example, it cannot easily support applications with continuous inputs or outputs, such as live audio transcoding or video conferencing.
Regardless of the chosen programming model, such applications will be particularly difficult to deploy on the LEO edge, given that service migration would have to occur seamlessly.
As these applications would benefit from low-latency edge deployment, we plan on investigating how they may be migrated efficiently in future work, possibly by integrating a continuous stream abstraction in a serverless platform.

\subsubsection*{Limitations of Replicated Data}

As in any replicated system, \komet{} is subject to data consistency tradeoffs~~\cite{abadi2012consistency} and end users inherit the consistency properties of the datastore used.

For applications with only single-writer access (either a single FaaS function instance or only one end user):
With the snapshot migration approach from §\ref{sec:migration}, end users will see consistent data for a non-replicated datastore but will encounter a period of downtime during the migration.
If such a datastore is replicated, end users will experience the consistency guarantees offered by that datastore, possibly enhanced with client-centric consistency guarantees~\cite{paper_bermbach2013_middleware_for_causal_consistency}.
In our approach, FReD~\cite{pfandzelter2023fred} already provides such client-centric guarantees by exposing vector clocks-based versioning and having the FReD client library request the appropriate versions from the datastore backend via the function instance used.
For the period of data migration, this means that our approach does not encounter downtime but will instead have small latency spikes when requests to data items that are not yet up-to-date on the new satellite are transparently served from the old replica instead.
We believe that this is preferable to downtime, especially considering that these not-too-high (it is a nearby satellite after all) latency spikes only occur when clients access data that has been written right before switching satellites.
All other data will already be up-to-date locally.

For all multi-writer scenarios:
The snapshot approach will have the downtime problem outlined above, in case of synchronous replication multiplied by the number of datastore replicas.
Furthermore, end users will again inherit the consistency and latency properties of the replication strategy used by the respective datastore -- from synchronous primary copy to asynchronous update everywhere.
In our approach, end users may not see updates from other end users instantly due to the asynchronous replication strategy used in FReD, i.e., they will encounter staleness, but client-centric consistency guarantees are provided.
Furthermore, there is no downtime and the mentioned latency spikes will happen at different points in time to different clients worldwide (whenever a migration happens), possibly with lower latency spikes as another replica might be closer than the one that is currently being migrated.

\subsubsection*{Satellite Server Selection}

We have shown in simulations that the scheduling heuristics we propose efficiently allocate satellite servers for services along the trade-off between service latency and migration frequency.
Nevertheless, the challenge of finding an optimal schedule requires far more research attention in the future.
Specifically, we plan to further investigate the costs of service latency and service migration.
Furthermore, there are additional factors that must be taken into account for service scheduling, such as satellite server resources, energy demand, or server temperature~\cite{bhosale4don,liu2024inorbit}.
The fluctuating demand for services, e.g., by bursty traffic, complicates scheduling further, requiring dynamic heuristics or an appropriate ever-changing solution to a formal assignment problem as services claim or release resources.
While fully solving distributed LEO edge scheduling is beyond the scope of this paper, we believe \komet{} makes meaningful progress toward this goal:
Its serverless architecture gives the platform fine-grained control over resource allocation at a per-request level, allowing it to adhere to any scheduling strategy.
The zero-downtime migration feature is especially valuable, enabling the platform to move services to nearby satellites with available resources.

\subsubsection*{Scheduler Location}

\komet{} includes a centralized, per-app\-li\-ca\-tion scheduler, which we believe is sufficient for our use-case, as scheduling decisions are not latency-critical given that they are not on the critical path of client requests.
More advanced distributed and fault-tolerant approaches are feasible yet outside the scope of this paper.
An interesting avenue for future research is the investigation of other scheduler architectures, such as a distributed, on-satellite scheduling component where each satellite server can make autonomous decisions, or a centralized scheduler that combines scheduling for all services.
For example, despite the possibility of per-application sharding (where each application is managed by a separate scheduler), scalability could become a challenge when handling many clients and potential satellite servers, particularly in many-to-one or many-to-many scenarios.
The main issue is distributing the heuristic calculations in a way that ensures they are completed within the required timeframes.
However, given that systems such as Starlink successfully use a similar centralized scheduling model, we are confident that such calculations are feasible.
Similarly, when considering fault tolerance, because scheduling occurs asynchronously and is not on a critical path, a scheduler failure would have minimal impact on LEO edge services.
A simple failover mechanism, such as using a shadow scheduler, would quickly restore normal operation.

\subsubsection*{Failure Tolerance}

LEO edge computing can be subject to on-board server failure due to radiation~\cite{pfandzelter2023failure,Bhattacherjee2020-kr}.
Although such failure is unlikely, the right failover mechanisms must be in place to provide continuous service coverage.
While this requires further research, we believe that the abstractions in \komet{} are suited for transparent failover.
Using data replication, a LEO edge service could keep a back-up secondary data replica on a nearby satellite at all times.
In case of satellite failure, the client could then be handed off to this secondary replica seamlessly, similarly to a normal hand-off.
While this comes with additional scheduling challenges and communication costs for additional replicas, it is only possible when services can be replicated easily by the underlying platform, as is the case in \komet{}.

\section{Conclusion}
\label{sec:conclusion}

This paper introduces \komet{}, a serverless platform for LEO edge computing.
\komet{} integrates FaaS compute abstractions and data replication to enable transparent service migration against satellite orbital movement to keep services deployed close to the clients that access them.
We implement a proof-of-concept prototype of \komet{} that we evaluate in an emulated testbed, demonstrating how the abstractions of \komet{} can be used to build three different example LEO edge applications.
Our evaluation shows that \komet{} can provide continuous service availability with high service levels despite satellite movement.
We also propose simple heuristics for service migration on the LEO edge that we evaluate based on our experiment traces.
Our simulation results show the trade-off between optimum service network distance and migration frequency and provide a starting point for future research on LEO edge service scheduling.

\begin{acks}
    We thank our anonymous reviewers and our shepherd Jon Crowcroft for their insightful feedback that helped shape this paper.
    We also thank Stanislav Kosorin for his support in evaluating container migration.
    This work is supported by the \grantsponsor{BMBF}{Bundesministerium für Bildung und Forschung (BMBF, German Federal Ministry of Education and Research)}{https://www.bmbf.de/bmbf/en} -- \grantnum{BMBF}{16KISK183} and \grantnum{BMBF}{01IS23068}.
\end{acks}

\bibliographystyle{ACM-Reference-Format}
\bibliography{bibliography}

\end{document}